\documentclass[conference]{IEEEtran}
\IEEEoverridecommandlockouts
\usepackage{cite}
\usepackage{amsmath,amssymb,amsfonts}
\usepackage{algorithmic}
\usepackage{graphicx}
\usepackage{textcomp}
\usepackage{xcolor}
\usepackage{tikz}
\usetikzlibrary{trees}
\usepackage{amsthm}
\usepackage{url}
\usepackage{enumitem}
\usepackage{booktabs}
\usepackage{multirow}
\usepackage{siunitx}
\begin{document}

\title{Opportunities and Challenges for Data Quality \\in the Era of Quantum Computing 
}

\author{\IEEEauthorblockN{1\textsuperscript{st} Sven Groppe}
\IEEEauthorblockA{\textit{Institute of Information Systems (IFIS)} \\
\textit{Universität zu Lübeck}\\
Lübeck, Germany \\
}
\and
\IEEEauthorblockN{2\textsuperscript{nd} Valter Uotila}
\IEEEauthorblockA{
\textit{University of Helsinki}\\
\textit{Aalto University}\\
Helsinki, Finland \\
}
\and
\IEEEauthorblockN{3\textsuperscript{rd} Jinghua Groppe}
\IEEEauthorblockA{\textit{Institute of Information Systems (IFIS)} \\
\textit{Universität zu Lübeck}\\
Lübeck, Germany \\
}
}

\maketitle

\begin{abstract}
In an era where data underpins decision-making across science, politics, and economics, ensuring high data quality is of paramount importance. Conventional computing algorithms for enhancing data quality, including anomaly detection, demand substantial computational resources, lengthy processing times, and extensive training datasets. This work aims to explore the potential advantages of quantum computing for enhancing data quality, with a particular focus on detection. We begin by examining quantum techniques that could replace key subroutines in conventional anomaly detection frameworks to mitigate their computational intensity. We then provide practical demonstrations of quantum-based anomaly detection methods, highlighting their capabilities. We present a technical implementation for detecting volatility regime changes in stock market data using quantum reservoir computing, which is a special type of quantum machine learning model. The experimental results indicate that quantum-based embeddings are a competitive alternative to classical ones in this particular example. Finally, we identify unresolved challenges and limitations in applying quantum computing to data quality tasks. Our findings open up new avenues for innovative research and commercial applications that aim to advance data quality through quantum technologies.
\end{abstract}


\section{Introduction}
The cornerstone of reliable decision-making in modern data-driven systems is data quality, which consists of dimensions such as accuracy, completeness, consistency, timeliness, and uniqueness \cite{Batini2006DQ}. However, ensuring high-quality data remains a computationally intensive task, particularly as datasets grow in volume, velocity, and variety. Traditional data quality tasks—such as data cleaning, deduplication, schema matching, and anomaly detection often require extensive processing power and time, especially when dealing with high-dimensional or noisy data. For instance, anomaly detection, which identifies outliers that deviate from expected patterns, involves complex statistical or machine learning algorithms that scale poorly with increasing data size, and often exhibit quadratic or higher computational complexity~\cite{chandola2009anomaly}. This complexity is exacerbated by the need to process real-time streams or massive historical datasets, and these render classical computing approaches inefficient for large-scale applications.

Quantum computing emerges as a promising paradigm to mitigate or even resolve some of these computational challenges. Built on the properties of quantum mechanics such as superposition, entanglement, interference, measurements and quantum parallelism, quantum algorithms offer - in comparison to algorithms designed for classical hardware - potential speedups for specific problems, for instance, unstructured search with quadratic speedups via Grover's algorithm~\cite{Grover1996} and linear algebra tasks like quantum principal component analysis with exponential speedups~\cite{lloyd2014quantum}. Table~\ref{tab:quantum_speedups} provides a further list of quantum algorithms with quadratic and exponential\footnote{Important caveats need to be considered that can limit the applicability of the quantum methods with exponential speedups~\cite{Aaronson2015FinePrint}.} speedups. The key enabler for many of the speedups is the HHL algorithm, which is a quantum algorithm for linear systems of equations~\cite{Harrow_Hassidim_Lloyd_2009,Wossnig_Zhao_Prakash_2018}. These capabilities could significantly reduce the time and resource demands of data quality tasks, particularly for problems involving high-dimensional optimization or pattern recognition. As quantum hardware with noisy intermediate-scale quantum (NISQ) devices of high capabilities becomes more accessible and fault-tolerant quantum computers are on the agenda of most quantum computer vendors until the end of this decade, the feasibility of applying quantum techniques to practical data management issues increases.

\begin{table}[htb]
\centering
\caption{Speedups of a selected list of quantum approaches}
\label{tab:quantum_speedups}
\begin{tabular}{p{1.5cm}p{6cm}}
\hline
\textbf{Speedup} & \textbf{Method and Reference} \\
\hline
Quadratic\newline $\mathcal{O}(\sqrt{N})$ & Bayesian inference~\cite{Low2014QBayesian,Wiebe2015SmallQuantum}\newline
Bayesian deep learning~\cite{Zhao_Pozas_Kerstjens_Rebentrost_Wittek_2019}\newline Online perceptron~\cite{Wiebe2016QuantumPerceptron}\newline Classical Boltzmann machine~\cite{Wiebe2014QDL}\newline Quantum reinforcement learning~\cite{Dunjko2016QEML}\newline Mean Estimation~\cite{Shyamsundar2021QMean}\newline Grammar-based Compression\cite{Gibney2023QGrammar}\newline kNN~\cite{Wiebe2014QkNNS}\newline
Gaussian process regression\cite{PhysRevA.99.052331}\\
Exponential\newline $\mathcal{O}(\log(N))$ & Least-squares fitting~\cite{Wiebe2012QDF}\newline Quantum Boltzmann machine~\cite{Amin2018QBM,Kieferov2017QBM}\newline Quantum PCA~\cite{Lloyd2014QPCA}\newline Quantum support vector machine~\cite{Rebentrost2014QSVM}\newline q-means~\cite{Lloyd2013QML}\newline kNN (in certain cases)~\cite{Wiebe2014QkNNS}\newline Fourier Transform~\cite{Asaka2020QFFT}\newline Wavelet transform~\cite{Bagherimehrab2024QWavelet}\newline Embeddings~\cite{Lloyd2020QE}\newline Evaluation of Distances~\cite{Lloyd2013QML}\newline
Decision Trees~\cite{Kumar2025DesQ}\\
\hline
\end{tabular}
\end{table}

Among data quality tasks, anomaly detection stands out as one of the most critical because it can address multiple quality dimensions simultaneously. Anomalies, whether they represent errors, fraud, or rare events, directly impact data accuracy since they introduce inconsistencies and lower reliability~\cite{aggarwal2017outlier}. Anomalies also affect completeness (missing expected patterns) and timeliness (delayed detection of deviations), and these make anomaly detection a linchpin for maintaining data integrity across domains like finance, healthcare, and cybersecurity. Classical methods, such as k-nearest neighbors, autoencoders, or statistical thresholding, often struggle with scalability and sensitivity in noisy or high-dimensional settings~\cite{Ghamry2024review}. Quantum-enhanced approaches, for instance, quantum support vector machines or quantum neural networks, have shown preliminary promise in accelerating outlier detection and improving robustness, particularly in hybrid quantum-classical frameworks~\cite{Corli2024ReviewQAD,guo2022quantum,Frehner2025qautoencoders,AparcanaTasayco2025QAD}. We argue that anomaly detection, as a computationally expensive yet essential data quality task, is an ideal candidate for quantum computing intervention, and its efficiency and effectiveness could potentially be revolutionized by quantum computing. 

Our main contributions are as follows:
\begin{itemize}
\item A study of quantum algorithms that can replace the computing-intensive subroutines in  classical anomaly detection approaches, to achieve potential benefits
\item A study of current quantum-based approaches to anomaly detection
\item An experimental study demonstrating the potential of quantum reservoir computing in detecting regime changes in time series data with the implementation publicly available on GitHub~\cite{valterUo_quantum_reservoir_regime_detection_2025}
\item Identifying open challenges and discussing future research directions
\end{itemize}

\section{Potential of Quantum Computing for\\Data Quality}

The integration of quantum computing into anomaly detection represents an underexplored yet promising frontier in enhancing data quality~\cite{Uotila2025QC4DQ}. 
A review of existing research shows that there is currently a lack of "true" quantum anomaly detection methods—whether methods inspired by classical approaches or their variations—indicating that purely quantum-native solutions have not yet been established. This gap highlights the need for novel quantum algorithms tailored specifically for anomaly detection tasks.

One potential direction is to develop quantum variants of established classical methods, such as the AutoRegressive Integrated Moving Average (ARIMA) model. Preliminary proposals suggest quantum replacements for ARIMA subroutines~\cite{Daskin2022QARIMA,Mujal2023QARIMA,Donatella2023QARIMA}, which could leverage quantum superposition and entanglement to enhance time-series anomaly detection. However, practical realizations remain speculative and require further investigation into quantum circuit design and error mitigation strategies.

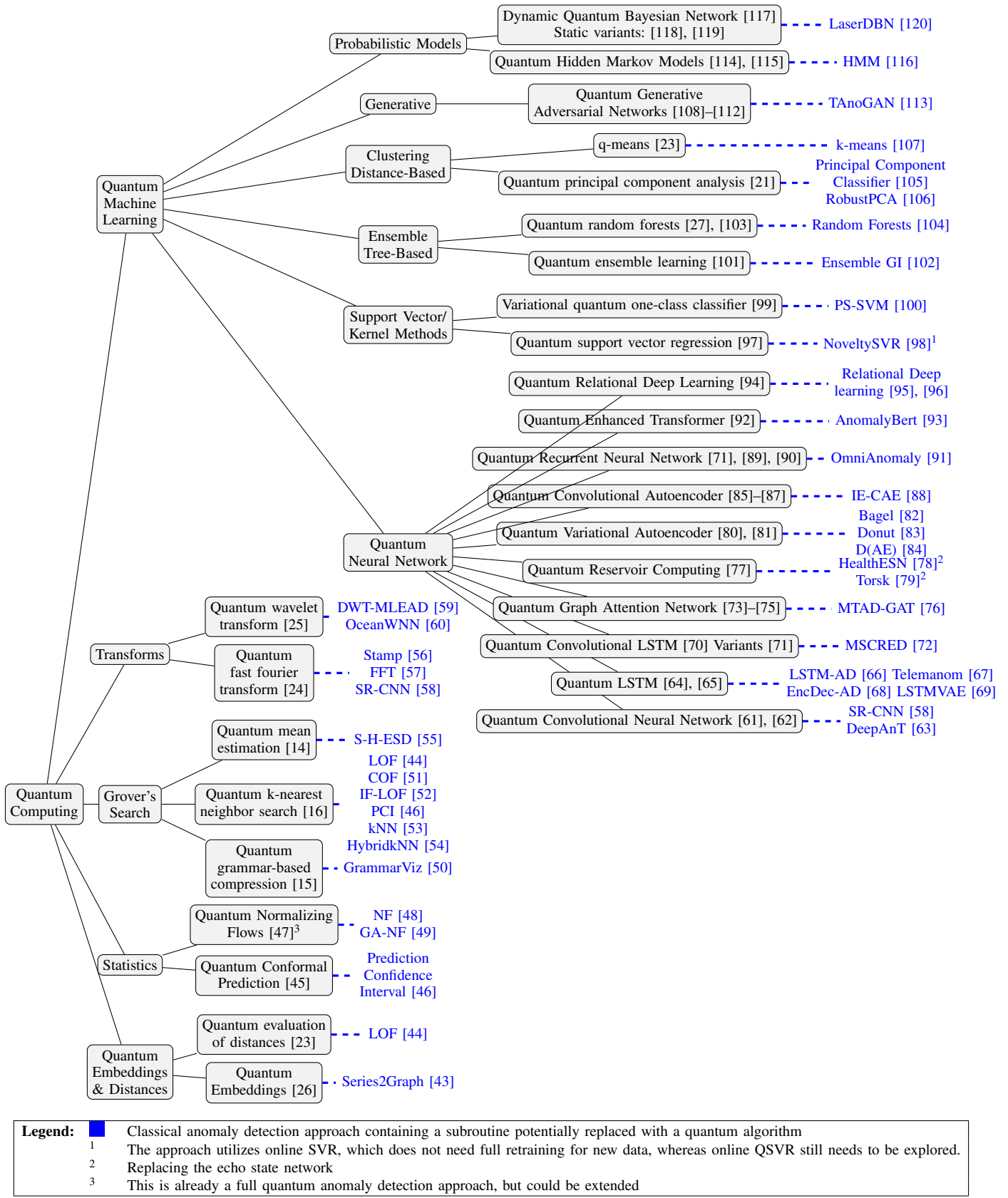
\begin{figure*}[htbp]
\textbf{Taxonomy of Quantum Approaches including ref. for potential drop-in replacements in\hfill {\color{blue} Classical Anomaly Detection}}\\
\noindent\makebox[\linewidth]{\rule{\textwidth}{0.4pt}}\vspace{0.4em}
\footnotesize
    \begin{tikzpicture}[align=center,grow=right,level distance=1cm,
  level 1/.style={sibling distance=5cm, level distance=1.6cm},
  level 2/.style={sibling distance=1.5cm,level distance=5cm},
  level 3/.style={sibling distance=0.7cm,level distance=4.5cm}]
  \node[rounded corners=.1cm,rectangle,draw,fill=gray!10]{Quantum\\Computing}
    child[sibling distance=2.5cm] {node[rounded corners=.1cm,rectangle,draw,fill=gray!10] {Quantum\\Embeddings\\\& Distances}
        child[sibling distance=0.4cm,level distance=2.5cm] {node[rounded corners=.1cm,rectangle,draw,fill=gray!10] {Quantum\\Embeddings \cite{Lloyd2020QE}}
          child[level distance=2.5cm,dashed,blue,very thick] { node{Series2Graph \cite{Boniol2020Series2Graph}}}
        }
        child[sibling distance=1.4cm,level distance=2.5cm] {node[rounded corners=.1cm,rectangle,draw,fill=gray!10] { Quantum evaluation\\of distances \cite{Lloyd2013QML}}
          child[level distance=2.5cm,dashed,blue,very thick] { node{LOF \cite{Breunig2000LOF}}}
        }
    }
    child[sibling distance=3cm] {node[rounded corners=.1cm,rectangle,draw,fill=gray!10] {Statistics}
        child[sibling distance=0.4cm,level distance=2.5cm] {node[rounded corners=.1cm,rectangle,draw,fill=gray!10] {Quantum Conformal\\Prediction \cite{Park2023QCP}}
          child[level distance=2.5cm,dashed,blue,very thick] { node{Prediction\\Confidence\\Interval \cite{Yu2014PCI}}}
        }
        child[level distance=2.5cm] {node[rounded corners=.1cm,rectangle,draw,fill=gray!10] {Quantum Normalizing\\Flows \cite{Rosenhahn2024QNF}\textsuperscript{3}}
          child[level distance=2.5cm,dashed,blue,very thick] { node{NF \cite{Ryzhikov2021NF}\\GA-NF \cite{dai2022graphaugmented}}}
        }
    }
    child {node[rounded corners=.1cm,rectangle,draw,fill=gray!10] {Grover's\\Search}
        child[sibling distance=1.2cm,level distance=2.5cm] {node[rounded corners=.1cm,rectangle,draw,fill=gray!10] {Quantum\\grammar-based\\compression \cite{Gibney2023QGrammar}}
          child[level distance=2.5cm,dashed,blue,very thick] { node{GrammarViz \cite{Senin2015GrammarVIZ}}}
        }
        child[sibling distance=1.2cm,level distance=2.5cm] {node[rounded corners=.1cm,rectangle,draw,fill=gray!10] {Quantum k-nearest\\neighbor search \cite{Wiebe2014QkNNS}}
          child[level distance=2.5cm,dashed,blue,very thick] { node{LOF \cite{Breunig2000LOF}\\COF \cite{Tang2002COF}\\IF-LOF \cite{Cheng2019IFLOF}\\PCI \cite{Yu2014PCI}\\kNN \cite{Ramaswamy2000kNN}\\HybridkNN \cite{Song2017HybridKNN}}}
        }
        child[sibling distance=1.2cm,level distance=2.5cm] {node[rounded corners=.1cm,rectangle,draw,fill=gray!10] {Quantum mean\\estimation \cite{Shyamsundar2021QMean}}
          child[level distance=2.5cm,dashed,blue,very thick] { node{S-H-ESD \cite{Hochenbaum2017SHESD}}}
        }
    }
    child[sibling distance=2.8cm] {node[rounded corners=.1cm,rectangle,draw,fill=gray!10] {Transforms}
        child[sibling distance=0.7cm,level distance=2.5cm] {node[rounded corners=.1cm,rectangle,draw,fill=gray!10] {Quantum\\fast fourier\\transform \cite{Asaka2020QFFT}}
          child[level distance=2.5cm,dashed,blue,very thick] { node{Stamp \cite{Yeh2016Stamp}\\FFT \cite{Rasheed2009FFT}\\SR-CNN \cite{Ren2019SRCNN}}}
        }
        child[sibling distance=1.4cm,level distance=2.5cm] {node[rounded corners=.1cm,rectangle,draw,fill=gray!10] {Quantum wavelet\\transform \cite{Bagherimehrab2024QWavelet}}
          child[level distance=2.5cm,dashed,blue,very thick] { node{DWT-MLEAD \cite{thill2017time}\\OceanWNN \cite{Wang2019OceanWNN}}}
        }   
    }
    child[sibling distance=5.6cm] {node[rounded corners=.1cm,rectangle,draw,fill=gray!10] {Quantum\\Machine\\Learning}
      child[sibling distance=2.6cm] {node[rounded corners=.1cm,rectangle,draw,fill=gray!10] {Quantum\\Neural Network}
          child {node[rounded corners=.1cm,rectangle,draw,fill=gray!10] {Quantum Convolutional Neural Network \cite{Cong2019QCNN,Yousif2024QCNN}}
              child[level distance=4.7cm,dashed,blue,very thick] { node{SR-CNN \cite{Ren2019SRCNN}\\DeepAnT \cite{Munir2019DeepAnT}}}
            }
          child {node[rounded corners=.1cm,rectangle,draw,fill=gray!10] {Quantum LSTM \cite{Chen2020QLSTM,Khan2024QLSTM}}
              child[level distance=4.7cm,dashed,blue,very thick] { node{LSTM-AD \cite{malhotra2015LSTMAD} Telemanom \cite{Hundman2018Telemanom}\\EncDec-AD \cite{Malhotra2016EncDecAD} LSTMVAE \cite{Park2018LSTMVAE}}}
            }
          child {node[rounded corners=.1cm,rectangle,draw,fill=gray!10] {Quantum Convolutional LSTM \cite{Xu2024QCLSTM} Variants \cite{Chen2025QRNN}}
              child[level distance=4.7cm,dashed,blue,very thick] { node{MSCRED \cite{Zhang2019MSCRED}}}
            }
          child {node[rounded corners=.1cm,rectangle,draw,fill=gray!10] {Quantum Graph Attention Network \cite{Ning2025QGAN,Yan2021QGAN,Liao2024QNN}}
              child[level distance=4.7cm,dashed,blue,very thick] { node{MTAD-GAT \cite{Zhao2020MTADGAT}}}
            }
          child {node[rounded corners=.1cm,rectangle,draw,fill=gray!10] {Quantum Reservoir Computing \cite{Fujii2020QRC}}
              child[level distance=4.7cm,dashed,blue,very thick] { node{HealthESN \cite{Chen2018HealthESN}\textsuperscript{2}\\Torsk \cite{Heim2019Torsk}\textsuperscript{2}}}
            }
          child {node[rounded corners=.1cm,rectangle,draw,fill=gray!10] {Quantum Variational Autoencoder \cite{Khoshaman2018QVA,Rao2023QVA}}
              child[level distance=4.7cm,dashed,blue,very thick] { node{Bagel \cite{Li2018Bagel}\\Donut \cite{Xu2018Donut}\\D(AE) \cite{Sakurada2014DAE}}}
            }
          child {node[rounded corners=.1cm,rectangle,draw,fill=gray!10] {Quantum Convolutional Autoencoder \cite{Kit2024QCA,Lo2025FER,Orduz2025QCA}}
              child[level distance=4.7cm,dashed,blue,very thick] { node{IE-CAE \cite{Garcia2021IECAE}}}
            }
          child {node[rounded corners=.1cm,rectangle,draw,fill=gray!10] {Quantum Recurrent Neural Network \cite{Bausch2020RQNN,Takaki2021VQRNN,Chen2025QRNN}}
              child[level distance=4.7cm,dashed,blue,very thick] { node{OmniAnomaly \cite{OmniAnomaly}}}
            }
          child {node[rounded corners=.1cm,rectangle,draw,fill=gray!10] {Quantum Enhanced Transformer \cite{prasanth2025quantumenhancedtransformer}}
              child[level distance=4.7cm,dashed,blue,very thick] { node{AnomalyBert \cite{Jeong2023AnomalyBert}}}
            }
          child {node[rounded corners=.1cm,rectangle,draw,fill=gray!10] {Quantum Relational Deep Learning \cite{Vogrin2024QRDL}}
              child[level distance=4.7cm,dashed,blue,very thick] { node{Relational Deep\\learning \cite{Fey2023RDL, Dwivedi2025RGT}}}
            }
        }
        child {node[rounded corners=.1cm,rectangle,draw,fill=gray!10] {Support Vector/\\Kernel Methods}
            child {node[rounded corners=.1cm,rectangle,draw,fill=gray!10] {   Quantum support vector regression \cite{Dalal2025QSVR}}
              child[dashed,blue,very thick] { node{NoveltySVR \cite{Ma2003NoveltySVR}\textsuperscript{1}}}
            }            
            child {node[rounded corners=.1cm,rectangle,draw,fill=gray!10] {   Variational quantum one-class classifier \cite{Park2023VQOCC}}
              child[dashed,blue,very thick] { node{PS-SVM \cite{Ma2003PS-SVM}}}
            }            
        }
        child {node[rounded corners=.1cm,rectangle,draw,fill=gray!10] {Ensemble\\Tree-Based}
            child {node[rounded corners=.1cm,rectangle,draw,fill=gray!10] {  Quantum ensemble learning \cite{Macaluso2020Ensemble}}
              child[dashed,blue,very thick] { node{Ensemble GI \cite{Gao2020Ensemble}}}
            }            
            child {node[rounded corners=.1cm,rectangle,draw,fill=gray!10] {  Quantum random forests \cite{Kumar2025DesQ,Srikumar2022QRF}}
              child[dashed,blue,very thick] { node{Random Forests \cite{Breiman2001RF}}}
            }        
        }
        child {node[rounded corners=.1cm,rectangle,draw,fill=gray!10] {Clustering\\Distance-Based}
            child {node[rounded corners=.1cm,rectangle,draw,fill=gray!10] { Quantum principal component analysis \cite{Lloyd2014QPCA}}
              child[dashed,blue,very thick] { node{Principal Component\\Classifier \cite{Shyu2003ANA}\\RobustPCA \cite{Paffenroth1018RobutsPCA}}}
            }
            child {node[rounded corners=.1cm,rectangle,draw,fill=gray!10] {q-means \cite{Lloyd2013QML}}
              child[dashed,blue,very thick] { node{k-means \cite{Yairi2001FaultDB}}}
            }
        }
        child[sibling distance=1.25cm] {node[rounded corners=.1cm,rectangle,draw,fill=gray!10] {Generative}
            child {node[rounded corners=.1cm,rectangle,draw,fill=gray!10] {Quantum Generative\\Adversarial Networks \cite{DallaireDemers2018,Lloyd2018,Hu2019,Huang2021GAN,Zoufal2019QGAN}}
              child[dashed,blue,very thick] { node{TAnoGAN \cite{Bashar2020TAnoGAN}}}
            }
        }
        child[sibling distance=1.2cm] {node[rounded corners=.1cm,rectangle,draw,fill=gray!10] {Probabilistic Models}
            child {node[rounded corners=.1cm,rectangle,draw,fill=gray!10] { Quantum Hidden Markov Models \cite{Markov2022QHMM,Srinivasan2017HQMM}}
              child[dashed,blue,very thick] { node{HMM \cite{Li2017HMM}}}
            }
            child {node[rounded corners=.1cm,rectangle,draw,fill=gray!10] { Dynamic Quantum Bayesian Network \cite{Borujeni2021DQBN}\\Static variants: \cite{Borujeni2021,Moreira2016}}
              child[dashed,blue,very thick] { node{LaserDBN \cite{Ogbechie2016}}}
            }
        }
    };
\end{tikzpicture}
\\\\\centering
\begin{tabular}{|@{}l@{}ll@{}|}
\hline
\hspace{0.5em}\textbf{Legend:}\hspace{1em} &
{\textcolor{blue}{\rule{1em}{1em}}} & Classical anomaly detection approach containing a subroutine potentially replaced with a quantum algorithm
\\
&\textsuperscript{1} & The approach utilizes online SVR, which does not need full retraining for new data, whereas online QSVR still needs to be explored.\hspace{0.5em} \\
& \textsuperscript{2} & Replacing the echo state network \\
& \textsuperscript{3} & This is already a full quantum anomaly detection approach, but could be extended \\\hline
\end{tabular}
\caption{%
Taxonomy of Quantum Computing algorithms that can serve as drop-in replacements of classical subroutines in the referenced approaches to anomaly detection.}
\label{fig:QuantumReplacementsInAD}
\end{figure*}

Another avenue lies in approaches utilizing quantum drop-in replacements for classical subroutines (see Figure~\ref{fig:QuantumReplacementsInAD}). For these hybrid methods, detailed in subsequent sections, we propose replacing components such as machine learning kernels or transformation algorithms with their quantum equivalents. For instance, quantum variants of Peaks-Over-Threshold (POT) methods, commonly used in algorithms like DSPOT~\cite{Siffer2017DSPOT}, have not yet been proposed despite related research~\cite{Manceau2019QPOT,Racs2023QPOT}. Such developments could extend quantum benefits to domains beyond anomaly detection and databases, potentially offering a higher impact if implemented successfully. The feasibility of these replacements, however, is hindered by unresolved challenges in practical implementation, including noise handling and scalability.

Conversely, certain classical approaches may not benefit from quantum enhancements due to their simplicity or the efficiency of existing classical implementations. Examples include MedianMethod \cite{Basu2006MedianMethod}, where quantum subroutines may introduce unnecessary complexity without performance gains. This underscores the importance of a context-specific evaluation to identify suitable candidates for quantum augmentation.

\begin{figure*}[htbp]
\textbf{Taxonomy of Quantum Approaches used in\hfill {\color{purple} Quantum Anomaly Detection}}\\
\noindent\makebox[\linewidth]{\rule{\textwidth}{0.4pt}}\vspace{0.4em}
\footnotesize
\begin{tikzpicture}[align=center,grow=right,level distance=1cm,
  level 1/.style={sibling distance=5cm, level distance=1.6cm},
  level 2/.style={sibling distance=1.5cm,level distance=5cm},
  level 3/.style={sibling distance=0.7cm,level distance=4.5cm}]
  \node[rounded corners=.1cm,rectangle,draw,fill=gray!10]{Quantum\\Computing}
    child[sibling distance=3cm] {node[rounded corners=.1cm,rectangle,draw,fill=gray!10] {Statistics}
        child[level distance=2.5cm] {node[rounded corners=.1cm,rectangle,draw,fill=gray!10] {Quantum Normalizing\\Flows \cite{Rosenhahn2024QNF}}
          child[level distance=2.5cm,dashed,purple,very thick] { node{\cite{Rosenhahn2024QNF}}}
        }
        child[level distance=2.5cm] {node[rounded corners=.1cm,rectangle,draw,fill=gray!10] {Quantum\\Density\\Estimation}
          child[level distance=2.5cm,dashed,purple,very thick] {node{\cite{liang2019quantum,Useche2024QDE}}}
        }
        child[level distance=2.5cm] {node[rounded corners=.1cm,rectangle,draw,fill=gray!10] {Quantum\\Amplitude\\Estimation}
          child[level distance=2.5cm,dashed,purple,very thick] {node{\cite{guo2022quantum,Guo2023QLOF,Pan2024QAD}}}
        }
    }
    child {node[rounded corners=.1cm,rectangle,draw,fill=gray!10] {Grover's\\Search}
        child[sibling distance=1.2cm,level distance=2.5cm] {node[rounded corners=.1cm,rectangle,draw,fill=gray!10] {Quantum k-nearest\\neighbor search \cite{Wiebe2014QkNNS}}
          child[level distance=2.5cm,dashed,purple,very thick] { node{\cite{Hdaib2024QAD,Kumar2025QAD,Pan2024QAD}}}
        }
    }
    child[sibling distance=12.6cm] {node[rounded corners=.1cm,rectangle,draw,fill=gray!10] {Quantum\\Machine\\Learning}
      child[sibling distance=2.6cm] {node[rounded corners=.1cm,rectangle,draw,fill=gray!10] {Quantum\\Neural Network}
          child {node[rounded corners=.1cm,rectangle,draw,fill=gray!10] {Quantum Convolutional Neural Network \cite{Cong2019QCNN,Yousif2024QCNN}}
              child[level distance=4.7cm,dashed,purple,very thick] { node{\cite{Kukliansky2024QAD,Amin2023QCNN,YasarArafath2023QCNN}}}
            }
          child {node[rounded corners=.1cm,rectangle,draw,fill=gray!10] {Tree Tensor Network \cite{Shi2006TTN}}
              child[level distance=4.7cm,dashed,purple,very thick] { node{\cite{Kukliansky2024QAD}}}
            }
          child {node[rounded corners=.1cm,rectangle,draw,fill=gray!10] {Multi-Scale Entanglement Renormalization Ansatz \cite{Vidal2008Mera}}
              child[level distance=4.7cm,dashed,purple,very thick] { node{\cite{Kukliansky2024QAD}}}
            }
          child {node[rounded corners=.1cm,rectangle,draw,fill=gray!10] {Quantum Variational Circuits}
              child[level distance=4.7cm,dashed,purple,very thick] { node{\cite{Wang2023VQC,Bhowmik24VQC,Wang2025VQC}}}
            }
          child {node[rounded corners=.1cm,rectangle,draw,fill=gray!10] {Quantum Support Vector Data Description}
              child[level distance=4.7cm,dashed,purple,very thick] { node{\cite{Oh2024QSVDD}}}
            }
          child {node[rounded corners=.1cm,rectangle,draw,fill=gray!10] {Quantum LSTM \cite{Chen2020QLSTM,Khan2024QLSTM}}
              child[level distance=4.7cm,dashed,purple,very thick] { node{\cite{Nguyen2024QLSTM}}}
            }
          child {node[rounded corners=.1cm,rectangle,draw,fill=gray!10] {Quantum Wavelet Neural Network}
              child[level distance=4.7cm,dashed,purple,very thick] { node{\cite{Huang2017QWNN}}}
            }
          child {node[rounded corners=.1cm,rectangle,draw,fill=gray!10] {Quantum Similarity Learning}
              child[level distance=4.7cm,dashed,purple,very thick] { node{\cite{Hammad2025SimilarityLearning}}}
            }
          child {node[rounded corners=.1cm,rectangle,draw,fill=gray!10] {Quantum Reservoir Computing \cite{Fujii2020QRC}}
              child[level distance=4.7cm,dashed,purple,very thick] { node{\cite{Ahmad2025QRC}}}
            }
          child {node[rounded corners=.1cm,rectangle,draw,fill=gray!10] {Quantum Autoencoder \cite{Khoshaman2018QVA,Rao2023QVA,Kit2024QCA,Lo2025FER,Orduz2025QCA}}
              child[level distance=4.7cm,dashed,purple,very thick] { node{\cite{Frehner2025qautoencoders,Hdaib2024QAD}\\\cite{ngairangbam2022anomaly,Sakhnenko2022Autoencoder}\\\cite{Ludmir25QAutoencoder,Kundu2025QAutoencoder}\\\cite{Araz24QAE,Ghosh2023QAE}}}
            }
          child[sibling distance=0.72cm] {node[rounded corners=.1cm,rectangle,draw,fill=gray!10] {Quantum Transfer Learning}
              child[level distance=4.7cm,dashed,purple,very thick] { node{\cite{Bhowmik24QTransferLearning}}}
            }
        }
        child {node[rounded corners=.1cm,rectangle,draw,fill=gray!10] {Support Vector/\\Kernel Methods}
            child[sibling distance=0.62cm] {node[rounded corners=.1cm,rectangle,draw,fill=gray!10] {Quantum support vector regression \cite{Dalal2025QSVR}}
              child[dashed,purple,very thick] { node{\cite{Hdaib2024QAD,Tscharke2023QSVR}\\\cite{Tscharke25QSVR}}}
            }            
            child {node[rounded corners=.1cm,rectangle,draw,fill=gray!10] {Unsupervised Quantum Kernel Machine}
              child[dashed,purple,very thick] { node{\cite{belis2024quantum}}}
            }            
            child[sibling distance=0.78cm] {node[rounded corners=.1cm,rectangle,draw,fill=gray!10] {Unsupervised one-class support vector machines}
              child[dashed,purple,very thick] { node{\cite{Kyriienko2022unsupervised,Pranjic23OCSVM}\\\cite{Cultice24QSVM,Tomono25QSVM}\\\cite{Kumar2025QAD,Chen25QSVM}\\\cite{Kolle23QSVM,Grossi2022QSVM}}}
            }            
        }
        child[sibling distance=1.15cm] {node[rounded corners=.1cm,rectangle,draw,fill=gray!10] {Quantum\\Boltzmann Machine}
            child {node[rounded corners=.1cm,rectangle,draw,fill=gray!10] {Quantum Restricted Boltzmann Machine}
              child[dashed,purple,very thick] { node{\cite{moro2023anomaly}}}
            }                    
            child {node[rounded corners=.1cm,rectangle,draw,fill=gray!10] {Quantum Boltzmann Machine}
              child[dashed,purple,very thick] { node{\cite{Stein2024QBMM}}}
            }                    
        }
        child {node[rounded corners=.1cm,rectangle,draw,fill=gray!10] {Ensemble\\Tree-Based}          
            child {node[rounded corners=.1cm,rectangle,draw,fill=gray!10] {  Quantum random forests \cite{Kumar2025DesQ,Srikumar2022QRF}}
              child[dashed,purple,very thick] { node{\cite{Hdaib2024QAD}}}
            }        
        }
        child[sibling distance=1cm] {node[rounded corners=.1cm,rectangle,draw,fill=gray!10] {Clustering\\Distance-Based}
            child {node[rounded corners=.1cm,rectangle,draw,fill=gray!10] {q-means \cite{Lloyd2013QML}}
              child[dashed,purple,very thick] { node{\cite{belis2024quantum}}}
            }
        }
        child[sibling distance=1cm] {node[rounded corners=.1cm,rectangle,draw,fill=gray!10] {Generative}
            child {node[rounded corners=.1cm,rectangle,draw,fill=gray!10] {Quantum Generative\\Adversarial Networks \cite{DallaireDemers2018,Lloyd2018,Hu2019,Huang2021GAN,Zoufal2019QGAN}}
              child[dashed,purple,very thick] { node{\cite{herr2021anomaly,Bermot2023QGAN}\\\cite{Kailasanathan25QGAN,Kalfon2024QGAN}}}
            }
        }
        child[sibling distance=1cm] {node[rounded corners=.1cm,rectangle,draw,fill=gray!10] {Probabilistic Models}
            child {node[rounded corners=.1cm,rectangle,draw,fill=gray!10] {Bayesian Quantum Orthogonal Neural Networks}
              child[dashed,purple,very thick] { node{\cite{Mathur25Bayesian}}}
            }
        }
    };
\end{tikzpicture}
\\\\\centering
\begin{tabular}{|@{}l@{}ll@{}|}
\hline
\hspace{0.5em}\textbf{Legend:}\hspace{1em} &
{\textcolor{purple}{\rule{1em}{1em}}} & Quantum anomaly detection approach with quantum subroutines
\hspace{0.5em} \\\hline
\end{tabular}
\caption{Taxonomy of the Quantum Computing algorithms, which have been used in quantum anomaly detection approaches}
\label{fig:QAD}
\end{figure*}
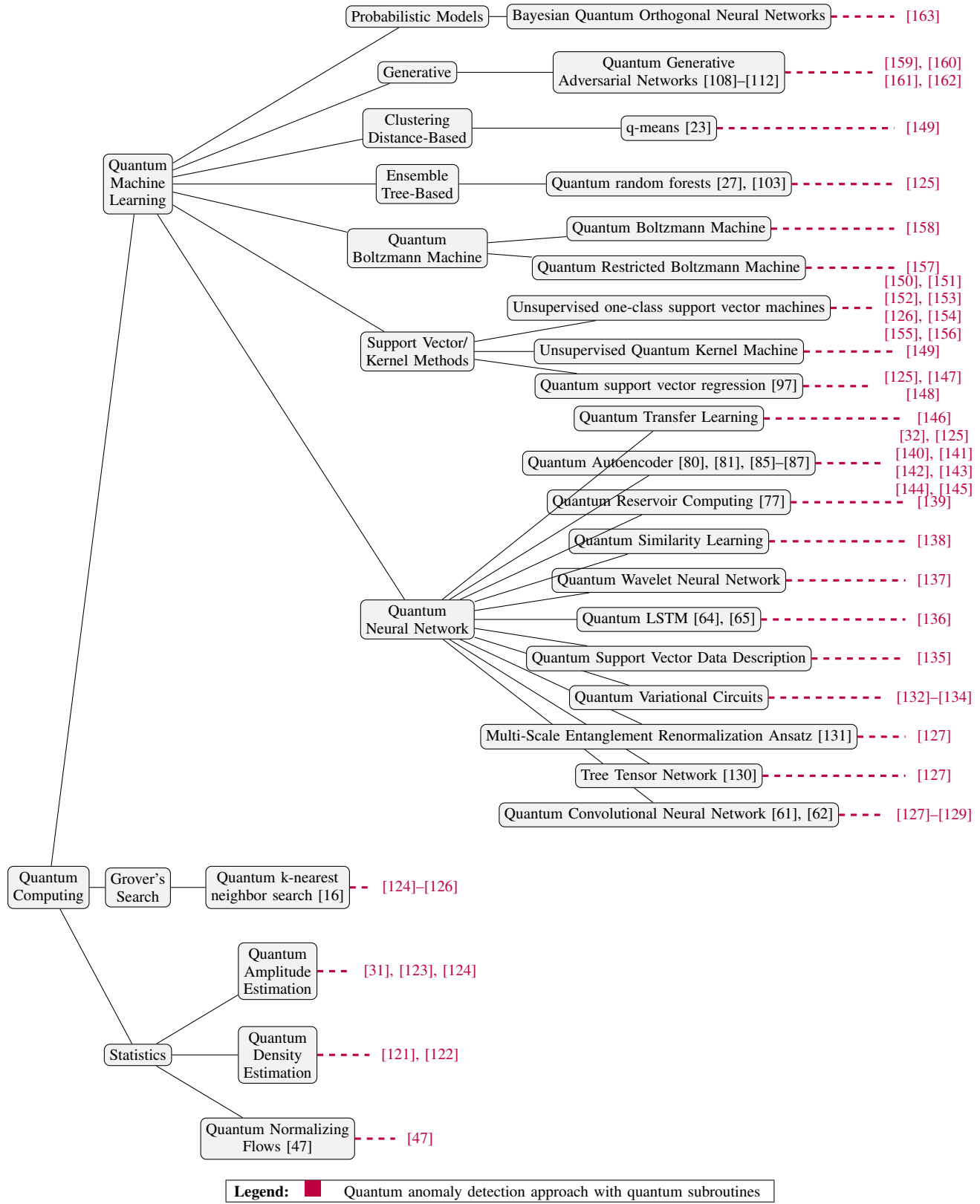

\subsection{Potential of Quantum Machine Learning} \label{sec:PotentialOfQML}

The potential of Quantum Machine Learning (QML) to enhance data quality offers a transformative paradigm for data-intensive applications. Preliminary investigations (as given e.g., in the referred quantum machine learning approaches in Figure~\ref{fig:QuantumReplacementsInAD}) suggest that QML can achieve performance levels comparable to, and in some cases superior to, classical machine learning on conventional hardware, with significantly fewer parameters - just $\approx$100 parameters compared to the thousands, millions, or even billions in classical models. This reduction in parameter count translates to a substantially lower memory footprint, making QML particularly appealing for resource-constrained environments.

QML enables more compact representations and requires only a logarithmic number of qubits in contrast to the linear scaling of classical bits. This property, coupled with potentially faster convergence, less training data, and enhanced generalizability, positions QML as a promising approach for modeling complex distributions that pose significant challenges for classical methods. The utilization of richer feature maps facilitated by quantum circuits offers a potential for improved separability and accuracy gains.
Intriguingly, noisy quantum gates may serve as a regularization technique \cite{Bagaev2025RegularizationByNoise,Domingo2023NoiseInQRC,Escudero2023ImpactOfNoise}, improving generalization by mitigating overfitting - a phenomenon that warrants further exploration.

The advantages of QML are, however, context-dependent. They are most pronounced with quantum-native data or when inputs can be efficiently prepared and encoded on quantum hardware. 
In classical benchmarking, its practical benefits are an active area of research, which is dedicated to addressing challenges such as data loading, noise management, and optimization issues. 

Today, quantum hardware still lacks support for a very high number of parameters in quantum machine learning architectures (see Section~\ref{sec:PotentialOfQML-QC}). With today's classical supercomputers, we also reach simulation limits of complex quantum models (see Section~\ref{sec:PotentialOfQML-Simulation}). Hence, the impact of employing a very high number of parameters in QML that could alter the observed performance trends remains uncharted so far. We discuss these and other issues, like availability of appropriate training data (see Section~\ref{sec:PotentialOfQML-Data}), that hinder investigating and achieving the full potential of quantum machine learning in practice, in the following paragraphs. 

\begin{figure}
  \includegraphics[width=\linewidth]{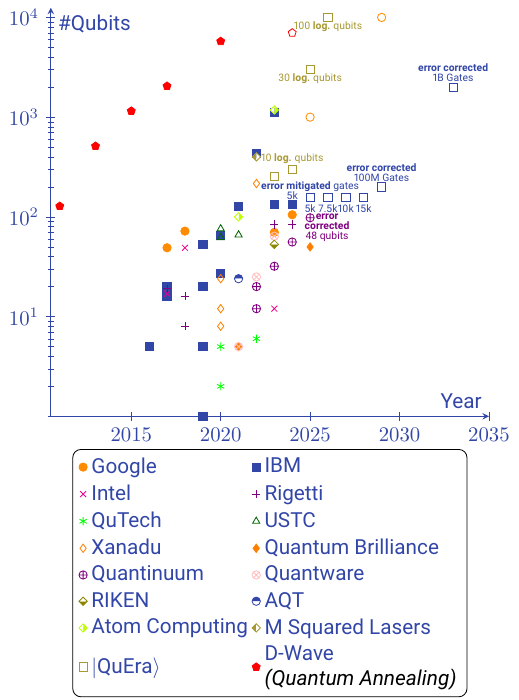}
  \caption{Timeline of the development of quantum computers across various vendors}
  \label{fig:QC}
\end{figure}

\subsubsection{Quantum Computer Wall}\label{sec:PotentialOfQML-QC}
The evolution of quantum computing hardware, as detailed in the Figure~\ref{fig:QC} provides hints for the rapid advancements and inherent limitations in the field. The growth in the number of qubits across leading quantum computing platforms - spanning big players like IBM and Google, but also many smaller companies like Rigetti, QuTech, Xanadu, Quantum Brilliance, Quantinuum, Atom Computing, QuEra, D-Wave, and many more - illustrates a trajectory to over a thousand qubits in the year 2025. Projections based on roadmaps from these quantum computer vendors suggest a continued increase, with error-mitigated systems reaching from 5,000 to 15,000 gates and error-corrected systems potentially scaling to 100 million or even 1 billion gates by the early 2030s.

The transition from error-mitigated to error-corrected quantum systems, anticipated around 2026–2028 with 10 to 100 logical qubits, represents a critical threshold for practical quantum advantage. However, current systems remain constrained by noise and error rates, necessitating advanced error correction techniques to harness the full potential of large-scale quantum computers. These developments emphasize the need for continued investment in hardware innovation to overcome the quantum computer wall and enable real-world applications in data quality enhancement.

\subsubsection{Simulation Wall}\label{sec:PotentialOfQML-Simulation}
The ``Simulation Wall'' delineates the computational boundaries of simulating quantum systems on classical hardware~\cite{Cicero2024QCSimulation}, a key consideration for validating quantum algorithms. Simulations storing all amplitudes for  30–34 qubits are feasible on single computing nodes or desktops with GPUs, and this corresponds to tens of gigabytes of memory. Large high-performance computing (HPC) clusters can perform simulations of 45–50 qubits 
\cite{Hner201745Qubits,Li201849Qubits,Raedt25Qubit50Simu}, which require hundreds of terabytes to a few petabytes. In general, when storing all amplitudes, the memory requirement doubles with each additional qubit. Hence, beyond 50 qubits, rendering a full simulation seems to be impractical with today's supercomputers due to astronomical resource demands.

However, approximations are possible for circuits with limited entanglement or shallow depth. For instance, reproducing Google's 53-qubit Sycamore approximate sampling experiment requires 512 GPUs over 15 hours \cite{Pan2022GoogleSimulation}, demonstrating the feasibility of hybrid approaches. These methods enable simulations of tens to over 1,000 qubits for specific tasks or circuits with reduced connectivity \cite{Patra2024SimulationOfOver1000Qubits}. This simulation wall poses a significant challenge for testing quantum anomaly detection and data quality algorithms, showing the necessity for advancing the development of real quantum computers. Future research should focus on optimizing hybrid simulation techniques to bridge the gap between classical simulation capabilities and the demands of large-scale quantum computations.

\subsubsection{Data (and Use Case) Wall}\label{sec:PotentialOfQML-Data}
The ``Data (and Use Case) Wall'' indicates critical data-related constraints in leveraging quantum machine learning (QML) for data quality improvement. 
The availability of high-quality data remains a bottleneck. Recent studies indicate that the data pool for training large language models is nearing exhaustion \cite{Villalobos2022EndOfData}, raising concerns about the scalability and further performance improvements of new machine learning architectures, including quantum variants, in general.

Noise within datasets can impede the prediction performance of learned models, such that there may exist inherent performance limits in terms of accuracy, mean squared error, etc., of a given use case. 
Neither classical machine learning nor quantum machine learning can surpass these performance limits. If classical machine learning methods have already reached these limits, the potential for quantum enhancement will be diminished.
As a result, a dual approach is required: improving data preprocessing to mitigate noise and conducting empirical studies to identify use cases where QML can offer a decisive advantage. This data-centric wall highlights the interplay between data availability, quality, and the practical applicability of quantum techniques.


\subsection{Potential of quantum computing in Anomaly Detection}


We are interested in classical subroutines in anomaly detection approaches, which require significant computing resources, because it makes sense to explore the potential of quantum computing in speeding up their processing. To this end, we performed two extensive literature reviews.

The first literature review studied which classical anomaly detection approaches could benefit from quantum computing. Two research results were achieved from the study: 
1) Identifying 44 classical anomaly detection approaches that could benefit
from using the corresponding quantum methods to replace their computing-intensive subroutines.
2) Developing a taxonomy of Quantum Computing algorithms that can serve as drop-in replacements of classical subroutines in the anomaly detection approaches identified above.
These research results are summarized in  Figure~\ref{fig:QuantumReplacementsInAD}. From this figure, we can also see that different types of anomaly detection techniques could potentially benefit from quantum subroutines.



Based on the results of the first literature review, we carried out the second one to investigate which quantum computing algorithms have been used in anomaly detection.
The works that use quantum computing have mostly developed their own anomaly detection approaches and do not directly adopt the efficient classical approaches by replacing their computing-intensive subroutines with quantum counterparts. Based on the study, we developed the second taxonomy, presented in Figure~\ref{fig:QAD}. 
  
From these two literature studies, we can see that although there have already been a significant number of quantum anomaly detection approaches proposed, there is still much space for future work, especially on developing quantum-based solutions that adopt efficient classical approaches but replace the time-consuming classical subroutines with quantum counterparts. 


\section{Experimental study on quantum reservoir computing}

In this experimental section, we demonstrate the usability of quantum computing by developing a quantum reservoir computing model, which is applied to a specific financial application: detecting volatility regime changes in time series market data~\cite{Hamilton_1989}. In other words, this means that the developed model aims to predict when the standard deviation of the stock returns exceeds a given threshold. The experiment is designed to be suitable for current noisy intermediate-scale quantum computers, and the tackled finance use case has a concrete, practical motivation. The model also avoids the most common issues in quantum machine learning, such as flat training landscapes and a high number of evaluations for a quantum computer. Since predicting how the standard deviation or other moments evolve over time in time series data is a general challenge, this method is likely to have applications beyond finance.

\subsection{Background on quantum reservoir computing}

Quantum reservoir computing is a special type of quantum machine learning paradigm, inspired by classical reservoir learning~\cite{10196105, https://doi.org/10.1002/qute.202100027}. The previous research has employed quantum reservoir computing for MNIST classification and Santa Fe laser data classification~\cite{kornjaca2024largescalequantumreservoirlearning}. This work is also closely related to the quantum reservoir learning use case in~\cite{Ahmad2025QRC}, which develops an intrusion and anomaly detection framework for edge-IoT systems. Our work and~\cite{Ahmad2025QRC} both develop a quantum reservoir learning model based on a similar type of quantum time evolution, measurements, and classical models. Both involve time series data with a window size. Our experimental implementation also differs from~\cite{Ahmad2025QRC} in several ways: we compare a wider range of classical models following the quantum reservoir, use a slightly more involved measurement scheme that includes computing pairwise Pauli-$Z$ expectation values, employ different Hamiltonians that encode the time series data, and address a different application domain. Our data preparation also demonstrates that the correspondence between the model's inputs and outputs is non-linear by construction, which is crucial for reservoir learning models. Before introducing quantum reservoir learning, we explain certain essential quantum mechanical concepts behind it.

Quantum computing is a computing paradigm that leverages the principles of quantum mechanics~\cite{Nielsen_Chuang_2010}. Quantum computation is typically a physical process in which a quantum mechanical system evolves, and at the end, the system is measured. The measured result describes something regarding the computational problem. The standard introduction to quantum computing usually involves defining qubits, quantum logical operations, and measurements. Especially, the quantum logical gates are often used to describe this time evolution. In this work, we do not explicitly require quantum logical gates, but the time evolution is described by Hamiltonian matrices $H$. Every quantum logical gate is generated by some Hamiltonian, so that in practice, it is sometimes more convenient to develop quantum algorithms in terms of Hamiltonians instead of gates. If $H$ is a Hamiltonian, then the unitary time evolution of the quantum mechanical system is described with
\begin{displaymath}
    U(t) := e^{-iHt},
\end{displaymath}
where $U$ is a time-dependent unitary matrix defined by Hamiltonian $H$. The key idea is that instead of defining quantum computations in terms of gates, one can define them in terms of Hamiltonians. 

Since the space of Hermitian operators forms a vector space, any Hamiltonian can be written as a linear combination of basis operators. The standard set of basis operators is the Pauli matrices $I$, $Z$, $X$, and $Y$~\cite{Nielsen_Chuang_2010}. While one can construct complex Hamiltonians with these Pauli matrices, this work focuses on the following three Hamiltonians
\begin{align}\label{eq:hamiltonians}
    H_X &= A_X\sum_{i=0}^{n} X_i, \\
    H_Z &= A_Z\sum_{i=0}^{n} x_i Z_i, \\
    H_{ZZ} &= A_{ZZ}\sum_{i=0}^{n-1} \left(x_i + x_{i+1}\right) Z_i Z_{i+1},
\end{align}
where $x_i$ for $i = 0, \ldots, n$ is a collection of datapoints that the Hamiltonians encode and $A_X$, $A_Z$, $A_{ZZ}$ are scalers. The full Hamiltonian is the sum $H = H_X + H_Z + H_{ZZ}$. We select these Hamiltonians based on the ones that were used in quantum reservoir computing in neutral atom quantum computers~\cite{kornjaca2024largescalequantumreservoirlearning}.

After the Hamiltonian evolution has been implemented using suitable Hamiltonians, the quantum mechanical system has to be measured so that we can access the outcome of the computation~\cite{uotila2025perspectives}. In this work, we employ a measurement scheme similar to that proposed in~\cite{kornjaca2024largescalequantumreservoirlearning}. The measurements are usually expressed in terms of measurement operators that, in this application, form a collection of single- and multi-qubit Pauli measurements in $Z$-basis as
\begin{equation}\label{eq:meas1}
\{\sigma^i_z \mid i = 1, \dots, n\} \text{ and } \{\sigma^i_z \sigma^j_z \mid 1 \le i < j \le n\}. 
\end{equation}

Now, the quantum reservoir computing follows the idea of classical reservoir computing. In the first phase, the input data is fed into the quantum reservoir, which in this case is encoded using Hamiltonians whose coefficients $x_i$ for $i = 0, \ldots, n$ represent the time series data. Then, the quantum mechanical system is measured, and the measurement results provide us with a vector that serves as an embedding in a higher-dimensional space. In the second phase, a classical, typically linear, model is fitted to the transformed data. If the classical model is fast and straightforward to compute, the reservoir training can be relatively fast.

\subsection{Volatility regime change detection}

The problem we want to tackle with quantum reservoir computing is detecting volatility regime changes in time series market data~\cite{Hamilton_1989}. While the problem is not commonly referred to as an anomaly detection problem, it shares many features with anomaly detection methods, such as the use of time series data and threshold-based decision criteria. We focus on 29 publicly traded stocks in the Dow Jones Industrial Average stock market index, spanning five years from 2015 to 2020. The data is obtained with yfinance~\cite{yfinance} from Yahoo! Finance's API. We then employed a standard financial data preparation process~\cite{Martin2021}. We have computed the expected returns $\left\{ p_t \mid t \in [t_{\mathrm{start}}, t_{\mathrm{end}}] \right\}$ and log-scaled their differences 
\begin{equation}\label{eq:returns}
R = \left\{ r_t = \ln(p_{t}/p_{t-1}) \mid t\in [t_{\mathrm{start}}, t_{\mathrm{end}} \right\}. 
\end{equation}
Next, we have fixed a window size $w > 0$. In this work, the window size coincides with the number of qubits, which is $9$. For this data, we have computed the standard deviation over the rolling window of size $w$ as
\begin{equation}\label{std_window}
    \sigma_{t}^{w} = \sqrt{\frac{1}{1 - w}\sum_{j = 0}^{w - 1}(r_{t-j} - \overline{r}_{t}^w)^2},
\end{equation}
where $\overline{r}_{t}^w = \frac{1}{w}\sum_{j=0}^{w - 1}r_{t-j}$ is the mean over the window. We have further normalized these values as
\begin{displaymath}
    \Tilde{\sigma}_t = \frac{\sigma_{t}^{w} - \mu_{\sigma}}{\sigma_{t}},
\end{displaymath}
where $\mu_{\sigma} = \frac{1}{|R| - w}\sum_{t = w}^{|R|}\sigma_{t}^{w}$ is mean over all rolling volatilities and
\begin{displaymath}
    \sigma_{t} = \sqrt{\frac{1}{|R| - w - 1}\sum_{t = w}^{|R|}(\sigma_t^{w} - \mu_{\sigma})^2}
\end{displaymath}
is the standard deviation of these rolling volatilities. We further compute the mean $\mu_{\Tilde{\sigma}}$ that describes the mean of the normalized volatilities and $\sigma_{\Tilde{\sigma}}$, which is the standard deviation of the normalized volatilities.

Then we concretely set the threshold, which we use to classify the volatility regimes, as $\tau := \mu_{\Tilde{\sigma}} + \lambda\sigma_{\Tilde{\sigma}}$, where we set $\lambda = 1$, but this can be adjusted depending on the investor's preferences. Then, we obtain a binary regime indicator as
\begin{displaymath}
    c_t = \begin{cases}
    1 \quad \text{if } \Tilde{\sigma}_t > \tau, \\
    0 \quad \text{otherwise}.
    \end{cases}
\end{displaymath}

As the previous calculations showed, the values $c_t$ depend on the return values non-linearly and thus, one can expect that a linear model would not perform well in this classification task. This problem can also be easily refined into a multi-class classification, where we would have more fine-grained regimes for volatility. We could also consider not only volatility but also higher moments, such as skewness and kurtosis, which would likely increase the complexity of the classification task. We consider these refinements to be part of future research, as well as employing this model on real hardware with an increased window size.

\subsection{Quantum reservoir learning model}

As a result of this data preparation, we obtain a binary classification problem that can be addressed using various supervised machine learning methods. The input data to the model is the log-scaled return values in Equation~\ref{eq:returns} grouped into chunks of size $w$, and the output values are the corresponding binary values $c_t$. For each batch $x_i$, for $i = 0, \ldots, w$, the corresponding values are substituted into the Hamiltonians in Equation~\ref{eq:hamiltonians}. The quantum circuits representing these Hamiltonians function as reservoirs, mapping feature vectors into higher-dimensional Hilbert spaces. As a result, we obtain a vector consisting of expectation values for measurement operators in Equation~\ref{eq:meas1}.

The quantum reservoirs are constructed using Qiskit, which is the most widely used quantum computing package maintained by the Qiskit community and IBM. Quantum circuits in the model are simulated with Qiskit without noise. The classical reservoirs are constructed with ReservoirPy~\cite{10.1007/978-3-030-61616-8_40}. Most of the classical models are imported from scikit-learn~\cite{scikit-learn} and PyTorch. We have utilized the grid search as a method for hyperparameter optimization. The implementation with the exact hyperparameter combinations is available on GitHub~\cite{valterUo_quantum_reservoir_regime_detection_2025}.

\subsection{Results}

For each case, we perform a hyperparameter search and then measure both accuracy and average precision scores. We used the first $80\%$ of the data for training and the rest $20\%$ was for testing. We also present the precision scores since the classification task is not well-balanced and the class $0$ is overrepresented. We average the accuracy and precision across the 29 stocks in the Dow Jones Industrial Average. The results on the test data set are presented in Table~\ref{table:results}, where we have selected the best-performing model from the grid search for each case.

\begin{table}[htbp]
\centering
\caption{Quantum reservoir vs classical reservoir vs raw}
\label{table:results}
\resizebox{\columnwidth}{!}{
\begin{tabular}{llSS}
\toprule
\textbf{Embedding} & \textbf{Model} & \textbf{Avg. accuracy} & \textbf{Avg. precision} \\
\midrule
\multirow{8}{*}{Quantum reservoir} 
 & Logistic Regression         & 0.923176 & 0.889995 \\
 & Support Vector Machine      & 0.920690 & 0.884758 \\
 & Ridge Classifier            & 0.912989 & 0.884359 \\
 & XGBoost                     & 0.895662 & 0.898514 \\
 & Random Forest               & 0.852056 & 0.849672 \\
 & Vanilla Neural Network      & 0.848851 & 0.793207 \\
 & Gaussian Mixture            & 0.760575 & 0.331688 \\
 & Multi-layer Perceptron      & 0.756552 & 0.668160 \\
\midrule
\multirow{3}{*}{Classical reservoir} 
 & Logistic Regression         & 0.895202 & 0.834589 \\
 & Ridge Classifier            & 0.894943 & 0.858850 \\
 & XGBoost                     & 0.843678 & 0.840786 \\
 & Support Vector Machine      & 0.830805 & 0.821675 \\
\midrule
\multirow{6}{*}{Raw, original data} 
 & Support Vector Machine      & 0.872069 & 0.903605 \\
 & XGBoost                     & 0.859080 & 0.843626 \\
 & Random Forest               & 0.823248 & 0.814772 \\
 & Vanilla Neural Network      & 0.769425 & 0.544674 \\
 & Logistic Regression         & 0.766437 & 0.472638 \\
 & Gaussian Mixture            & 0.760204 & 0.505326 \\
 & Multi-layer Perceptron      & 0.758966 & 0.461540 \\
 & Ridge Classifier            & 0.758391 & 0.473946 \\
\bottomrule
\end{tabular}
}
\end{table}


These results demonstrate that quantum reservoir computing performs at least as well as classical reservoir computing, outperforming classical linear models in the task and also performing well against non-linear classical models, such as a support vector machine with a non-linear kernel. Considering that logistic regression and ridge classifiers are simple and fast to train, quantum reservoir computing may be a viable approach for pipelines where constant retraining and model adaptation are necessary. Additionally, we observe that applying reservoirs clearly improves the performance of the linear models in both quantum and classical reservoir cases. The initial results, which are comparable to those obtained with classical methods, are positive and encouraging, especially considering that this is preliminary work and we used only $9$ qubits. These results will motivate further studies of quantum reservoir-based models in tasks where the learning task involves non-linearities.

\section{Open Challenges and Future Research Directions}

The integration of quantum computing into data quality management opens a frontier of opportunities and challenges. While preliminary investigations suggest significant potential—particularly for anomaly detection—extending these benefits to other data quality tasks such as data cleaning, deduplication, data imputation, and schema matching requires concerted research efforts. This section discusses future research directions and identifies the open challenges that must be overcome to realize the full promise of quantum-enhanced data quality solutions.

A primary direction is the development of quantum-native algorithms tailored to data quality tasks. Future work should focus on designing end-to-end quantum anomaly detectors that leverage quantum superposition and entanglement, potentially achieving exponential speedups in outlier identification. Beyond anomaly detection, quantum algorithms could address data cleaning by optimizing error correction in noisy datasets using quantum error mitigation techniques~\cite{Zhang2025ErrorMitigation}, or enhance deduplication through quantum clustering methods like q-means~\cite{Lloyd2013QML}. Data imputation may utilize a similar set of quantum approaches to anomaly detection. Schema matching, which involves aligning heterogeneous data structures, could benefit from quantum graph neural networks to model complex relationships more efficiently \cite{verdon2019quantum}.

Another promising avenue is the experimental validation of hybrid quantum-classical systems on NISQ devices. As quantum hardware evolves, pilot studies should evaluate the performance of quantum subroutines (e.g., quantum fast Fourier transforms for time-series anomalies) against classical baselines across diverse datasets. Additionally, the development of quantum software frameworks tailored for data quality tasks, integrating with existing platforms like Apache Spark, could accelerate adoption. Research into quantum data encoding strategies is also critical, ensuring efficient loading of classical data onto quantum hardware to mitigate the ''data-loading bottleneck''~\cite{Biamonte2017QML}.

Despite these opportunities, several challenges remain. First, the current limitations of NISQ devices (as described in Section~\ref{sec:PotentialOfQML-QC}), characterized by high error rates and limited qubit coherence, pose significant barriers. Noise mitigation strategies, such as those proposed in~\cite{Zhang2025ErrorMitigation}, need refinement to ensure robustness in data quality applications. Second, the ``Simulation Wall'' (as described in Section~\ref{sec:PotentialOfQML-Simulation}) restricts the ability to test quantum algorithms beyond 50–60 qubits on classical hardware, necessitating access to scalable quantum infrastructure. Third, the ''Data Wall'' (as described in Section~\ref{sec:PotentialOfQML-Data}) highlights the scarcity of high-quality training data and the impact of noise within datasets, which could undermine quantum advantages unless addressed through advanced preprocessing.

Specific to anomaly detection, the practical realization of quantum drop-in replacements is often hindered by unresolved issues in online learning and scalability. For data cleaning, the computational cost of quantum error correction may outweigh the benefits for simple datasets, requiring a cost-benefit analysis. Deduplication faces challenges in adapting quantum search algorithms (e.g., Grover's search~\cite{Grover1996}) to efficiently handle approximate string matching. Schema matching, meanwhile, demands robust quantum embeddings to capture semantic similarities, an area where current quantum graph methods are still immature~\cite{Lloyd2020QE}.

\section{Summary and Conclusions}

In this visionary work, we study the potential of quantum computing for data quality. We first performed extensive literature studies and generated two taxonomies: One discovers a drop-in replacement categorization between classical and quantum subroutines, especially focusing on anomaly detection; the other provides a landscape of current research work on quantum-based solutions to anomaly detection. The taxonomies show that quantum computing presents a vast set of possibilities for exploring the potential benefits of quantum computing in data quality tasks. To concretely demonstrate the possibilities, we carried out an experimental evaluation of quantum reservoir computing for detecting regime changes in stock market data. The evaluation results indicated that quantum reservoir learning was a competitive alternative to the classical models in this specific task.

Ultimately, the interdisciplinary nature of this field necessitates collaboration among quantum physicists, data scientists, and database engineers to bridge theoretical advancements with practical applications. Ethical considerations, such as the energy consumption of quantum systems versus classical alternatives, also warrant investigation. Addressing these challenges through targeted research and international initiatives will be essential to unlocking the transformative potential of quantum computing for data quality tasks.

\bibliographystyle{IEEEtran}
\bibliography{bibfile}

\end{document}